\documentclass[pre,aps,twocolumn,showpacs,preprintnumbers,amsmath,amssymb,superscriptaddress]{revtex4}

\usepackage{graphicx}
\usepackage[ansinew]{inputenc}
\usepackage{epsfig}
\begin{document}
\title{Logarithmic bred vectors in spatiotemporal chaos: structure and growth}
\author{Sarah Hallerberg}
\email{sarah.hallerberg@physik.tu-chemnitz.de}
\affiliation{Instituto de F{\'i}sica de Cantabria (IFCA), CSIC-Universidad de
Cantabria, E-39005 Santander, Spain}
\affiliation{Institute of Physics, Chemnitz University of Technology, D-09107
Chemnitz, Germany}
\author{Diego Paz{\'o}}
\email{pazo@ifca.unican.es}
\affiliation{Instituto de F{\'i}sica de Cantabria (IFCA), CSIC-Universidad de
Cantabria, E-39005 Santander, Spain}
\author{Juan M. L{\'o}pez}
\email{lopez@ifca.unican.es}
\affiliation{Instituto de F{\'i}sica de Cantabria (IFCA), CSIC-Universidad de
Cantabria, E-39005 Santander, Spain}
\author{Miguel A. Rodr{\'i}guez}
\email{rodrigma@ifca.unican.es}
\affiliation{Instituto de F{\'i}sica de Cantabria (IFCA), CSIC-Universidad de
Cantabria, E-39005 Santander, Spain}

\date{\today}

\begin{abstract}
Bred vectors are a type of finite perturbation used in prediction studies of
atmospheric models that exhibit spatially extended chaos. 
We study the structure, spatial correlations, and the growth-rates of
logarithmic bred vectors (which are constructed by using a given norm).
We find that, after a suitable transformation,
logarithmic bred vectors are roughly piecewise copies of the leading Lyapunov
vector.
This fact allows us to deduce a scaling law
for the bred vector growth rate as a function of their amplitude.
In addition, we relate growth rates with the spectrum
of Lyapunov exponents corresponding to the most
expanding directions. 
We illustrate our results with simulations of the Lorenz '96 model.
\end{abstract}

\maketitle
%
\section{Introduction}
In dynamical systems theory, chaos is usually defined on the basis of
the exponential departure of infinitesimally separated initial conditions.
The exponential law stems from the linearity of the equations that govern
infinitesimal perturbations.
In applications, however, errors are typically {\em finite}. 
This occurs, for instance, in the breeding method used to
generate finite perturbations for ensemble forecasting 
at the National Centers for environmental Prediction
(USA)~\cite{TothKalnay1993,kalnay}.
Bred vectors are 
closely related to Lyapunov vectors (LVs) albeit different since they are
finite by construction and result from the evolution of perturbations,
which are imposed to have a certain size via periodic normalizations.
One advantage of bred vectors in applications is that the model
under study does not have to be linearized, in contrast with LVs. 
(Keep in mind that the linearization of a meteorological model is a 
delicate question
due to the presence of nondifferentiable or discrete programming structures, see e.g.~\cite{tlm}.)
The drawback is that bred vectors are governed by fully nonlinear models,
what constitutes a challenge for their theoretical description.

In this article we use a special class of bred vector, the so-called
{\em logarithmic bred vector} (Log-BV)~\cite{Primo2005,Primo2006}, which has the
particularity of being
normalized using the geometric (or zero) norm. 
This choice has proven to be the most convenient
for spatially extended systems (see Ref.~\cite{Primo2005,Primo2006}) and
offers some advantages in theoretical terms.
We will focus on two aspects of bred vectors:
First, we uncover common structural properties of Log-BVs
and the leading LV. In the second part of the 
paper we study the growth rate of Log-BVs, which we shall
refer to as {\em bred exponents} (BEs)
---instead of the more imprecise term finite-size Lyapunov
exponents~\cite{Boffetta2002}---
for the sake of brevity and to emphasize their origin. 
We find an interesting relation between
BEs and Lyapunov exponents (LEs) of spatio-temporally chaotic systems.

This paper is organized as follows: In Sec.~\ref{lorenz96} we introduce the
Lorenz '96 model, which we will use throughout this contribution in order to
illustrate our considerations. For the parameter values selected
the Lorenz '96 is hyperchaotic and exhibits spatio-temporal chaos.
In the following section we explain the computation of Log-BVs.
A surface picture is introduced for the leading
LV in Sec.~\ref{sgfa} and extended to analyze Log-BVs
in Sec.~\ref{sgfb}.
In Sec.~\ref{gr} we study the growth rates ({\em i.e.}, BEs)
of Log-BVs.
In Sec.~\ref{sr} the convergence to the largest LE is found to obey
a scaling relation.
In Sec.~\ref{non-leading} we explain how the BEs
are related to the spectrum of LEs.
The conclusions are summarized in Sec.~\ref{conc}.

%
\section{The Lorenz '96 model}
\label{lorenz96}
The Lorenz '96 model~\cite{Lorenz1996, Lorenz1998} is a toy model
originally proposed in the
context of atmospheric dynamics and used extensively to test novel techniques
and applications. 
It is a time continuous model consisting of a set of nonlinear ODEs 
coupled in a ring geometry:
\begin{eqnarray}
\frac{d \,u(x,t)}{dt}  &=& -u(x,t) - u(x-1,t)u(x-2,t)\nonumber\\
&\it{}&\quad +u(x-1,t)u(x+1,t) +F,\quad \nonumber\\
&\it{}& \mbox{with}\quad x=1,...,L.
\label{lorenz}
\end{eqnarray}
$u(x,t)$ can be seen as a scalar meteorological variable, 
{\em e.g.}~temperature, at $L$ equally spaced sites $x$ on a latitude circle
(and hence periodic
boundary conditions are assumed).
Moreover, there is an external forcing constant $F$ that mimics
the solar driving of the atmosphere.

For $L>3$ the solutions of (\ref{lorenz}) are chaotic if $F$ is large enough. In
particular 
the steady solution $u(x,t)=F$ becomes unstable if $F >8/9$ \cite{Lorenz1998}. A
more detailed
study shows that stable nonchaotic solutions survive up to a value of $F$ that,
though depending on $L$,
is approximately in the range 4 to 6 \cite{Lorenz2006}. 
Beyond some threshold value of $F$, chaotic 
dynamics of the model becomes fully developed. More precisely, the dynamics is
extensive with $L$ (see e.g. results
in \cite{Karimi2009} for $F=10$).
 Extensivity means that many relevant quantities
(dimension, entropy, etc.) scale linearly with the system size, and the LEs
$\{\lambda_n\}_{n=1,\ldots,L}$ converge to a density in the ``thermodynamic''
limit $L\to \infty$.
This property is shared by a number of extended dynamical
systems ranging from coupled map lattices to PDEs such as the
Kuramoto-Sivahinsky or the complex Ginzburg-Landau
equations \cite{CrossHohenberg}.

In order to compute bred and Lyapunov vectors and their growth rates in the
Lorenz '96 model, we
integrate Eq.~(\ref{lorenz}) and its linearization (tangent space) by using a
fourth order
Runge-Kutta-solver with  time step  $\Delta t=0.01$.
Before measuring the quantities we are interested in, 
we allow the system to breed Log-BVs for a transient time
$t_{trans} > 8(L^{1.5})$.
%
\section{Logarithmic Bred Vectors and Breeding}
\label{breeding}
In 1993, Toth and Kalnay \cite{TothKalnay1993} created a special
operational cycle designed to ``breed'' fast growing errors
in meteorological models. It uses finite perturbations
that are periodically normalized to become bred vectors after some breeding
cycles.

In the following we describe the procedure to compute bred vectors 
in mathematical terms.
A control (unperturbed) trajectory $\mathbf{u}(t)=[u(x,t)]_{x=1}^{x=L}$
and a perturbed one $\tilde{\mathbf{u}}(t)=[\tilde{u}(x,t)]_{x=1}^{x=L}$ are 
evolved in parallel obeying Eq.~(\ref{lorenz}).
The difference between the perturbed and unperturbed trajectories is calculated  
every time interval $\tau$, say at times $\tau_m = m \tau$, with $m =1, 2,
\ldots$ to obtain
\begin{equation}
\delta{\mathbf{u}}(\tau_m) = \tilde{\mathbf{u}}(\tau_m) - \mathbf{u}(\tau_m).
\end{equation}
These differences are then scaled down at $t=\tau_m$ to a given perturbation
amplitude $\varepsilon_0$ by defining
\begin{equation}
\delta \mathbf{u}_{\varepsilon_0}(\tau_m) = \varepsilon_{0} \frac{\delta
\mathbf{u}(\tau_m)}{ \|\delta \mathbf{u}(\tau_m) \|},
\end{equation}
where $\delta \mathbf{u}_{\varepsilon_0}(x,\tau_m)$ denotes the bred vector at
time $\tau_m$, and is $\| \cdot \|$ a particular norm.
The perturbed trajectory is then redefined by means of the bred vector:
\begin{equation}
\tilde{\mathbf{u}}(\tau_m^+)= \mathbf{u}(\tau_m) + \delta
\mathbf{u}_{\varepsilon_0}(\tau_m),
\end{equation}
with $\tau_m^{+}=\lim_{\nu \rightarrow 0} \tau_m + \nu$ referring to the
same time $\tau_m$, but after the rescaling. 
Perturbed and unperturbed trajectories are again integrated forward
in parallel until next rescaling scheduled at time $\tau_{m+1}$.

The definition of the bred vector contains two ingredients. One is the parameter
$\varepsilon_0$ controlling the amplitude. 
The second ingredient is the norm to be used since different norms will produce
different bred vectors.
In recent works \cite{Primo2005,Primo2006} it was found that the $0$-norm
(or geometric norm)
\begin{equation}
 \| \delta \mathbf{u}(t) \| \equiv \lim_{q\to0}\left[\sum_{x=1}^{L} |\delta
u(x,t)|^q\right]^{1/q} = \prod_{x=1}^{L} |\delta u (x,t)|^{1/L},
\end{equation}
is a convenient choice for breeding. 
Due to the multiplicative character
of the linear dynamics the $0$-norm ought to produce bred vectors that, at
different times,
$\tau_m$ are the most statistically equivalent among them. 
The bred vectors constructed in this way are called {\em logarithmic} bred
vectors (log-BVs) \cite{Primo2005,Primo2006,Pazo2010}.
%

\section{Structure of logarithmic bred vectors}

\subsection{The main Lyapunov vector}
\label{sgfa}
Before starting our analysis of the Log-BVs it is useful to recall
recent results concerning the structure of the main (or leading)  LV.
In a dynamical system an infinitesimal perturbation evolves generically towards
the leading Lyapunov vector $\mathbf{g}(t)$,
which indicates the direction of maximal growth for perturbations integrated
since the remote past.
The orientation of the LV in tangent space 
depends on the position in the chaotic
attractor (that can be parametrized by time).
For extended chaotic systems, it is observed that
the LV projects very inhomogeneously on space.
More precisely, the vector localizes at some quite narrow region of the
system \cite{GiacomelliPoliti1991}. But noticeably, this localization is dynamic
and the localization center changes as time evolves, thus recovering the
homogeneity of the system in a statistical sense.
In the 90s Pikovsky and coworkers \cite{Pikovsky1994,PikovskyPoliti1998} found
very useful for the theoretical analysis
to associate a ``surface'' with the leading LV by means of a
Hopf-Cole transformation,
which in $d=1$ dimension reads:
\begin{equation}
h_{LV}(x,t) = \ln |g(x,t)|,  \label{Lyapuhopfcole}
\end{equation}
with ${\mathbf g}(t) =[g(x,t)]_{x=1}^{x=L}$. 
For a large family of systems \cite{PikovskyPoliti1998},
including the Lorenz '96 model~\cite{Pazo2008}, $h_{LV}(x,t)$ exhibits
correlations in space and time which are described by the canonical
Kardar-Parisi-Zhang (KPZ) equation of stochastic surface growth \cite{Kardar1986}.
This mapping leads to interesting scaling properties for $h_{LV}(x,t)$. 
The average width $w= \langle (h_{LV}-\langle h_{LV}
\rangle)^2\rangle^{1/2}$ scales
with the length of the system \cite{Barabasi1995, Halpin-Healy1995}
as $w\propto L^{\alpha}$; with $\alpha=1/2$
as in the KPZ equation in one dimension. This means that, at sufficiently long
scales, $h_{LV}(x,t)$ appears as the path of a random walk in $d=1$.
This self-affine profile translates into a power-law dependence of
the  structure factor (a spatial power spectrum) at small wavenumbers:
\begin{equation}
S(k) \propto k^{-(2\alpha+d)} \; \; {\rm for} \; \; k<\bar k,
\end{equation}
where $S(k)= \langle \hat{h}(k,t) \hat{h}(-k,t)\rangle_t$,
and $\hat{h}(k,t) = \sum_x \exp(ikx) h(x,t)$.
At short length scales, $k > \bar k$, non-universal short-range correlations
are expected to appear due to the deterministic character of the system.
However, below $\bar k$ universal scaling properties emerge and this is
reflected in a spatial correlation that generically decays as $\sim k^{-2}$ in
one dimension.
\begin{figure}
\centerline{
\epsfig{file=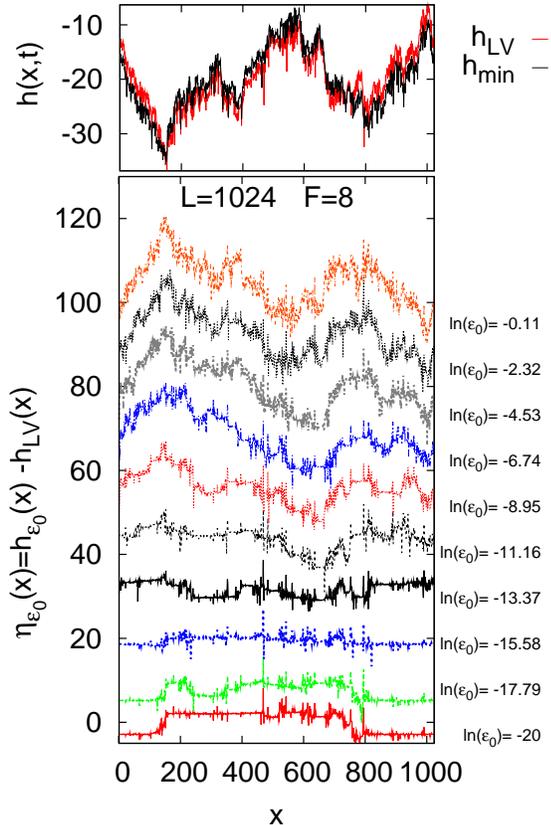,width=8cm}}
\caption{\label{difffields} 
(color online) Upper panel: Surfaces associated with the LV and with the
Log-BV with the smallest value of $\varepsilon_0$ used $\varepsilon_0^{min}=e^{-20}$
(Lorenz '96 model with $F=8.0$ and $L=1024$). 
Lower panel: Snapshot of the difference fields $\eta_{\varepsilon_0}= h_{\varepsilon_0}(x,t) -
h_{LV}(x,t)$ (curves are arbitrarily shifted to improve their visibility).}
\label{plateaus}
\end{figure}

\subsection{The logarithmic bred vectors}
\label{sgfb}
Analogously to the surface associated with the leading LV in
Eq.~(\ref{Lyapuhopfcole}), one can define a surface 
\begin{equation}
h_{\varepsilon_0}(x,t)=\ln |\delta u_{\varepsilon_0}(x,t)| 
\label{surflbv}
\end{equation}
by Hopf-Cole transforming the Log-BV of amplitude $\varepsilon_0$.

The main LV evolves following linear equations and, therefore,
only its direction in tangent space matters.
However for a bred vector the norm plays a very important role through the periodic
rescaling of the perturbations defined above. If the geometric norm is chosen,
the spatial mean of the associated Log-BV surface is fixed to a given size
$\varepsilon_0$:
\begin{equation}
\overline{h}=\frac{1}{L} \sum_{x=1}^{L} h_{\varepsilon_0}(x,t=\tau_m) =\ln \varepsilon_0.
\label{defhm}
\end{equation}
In the limit of small amplitude $\varepsilon_0 \to 0$, the Log-BV aligns
with the leading Lyapunov vector,
and their profiles coincide, $h_{\varepsilon_0\to 0}(x,t)=h_{LV}(x,t)$, 
apart from an arbitrary constant due to the arbitrary norm
(i.e.~arbitrary $\bar{h}$) of the leading LV. 
This fact suggests that the leading LV can be a good reference
point to analyze the structure of Log-BVs.
In Fig.~\ref{difffields} we show a snapshot of the
difference fields 
\begin{equation}
\eta_{\varepsilon_0}(x,t)= h_{\varepsilon_0}(x,t) - h_{LV}(x,t).
\end{equation} 
We can see that, unless $\varepsilon_0$ is too large,
$\eta_{\varepsilon_0}$ is mainly
composed of plateaus, which indicates that
the structure of $h_{\varepsilon_0}(x,t)$ is roughly a piecewise copy of
$h_{LV}(x,t)$.
Remarkably, this structure has been previously observed for LV-surfaces associated
with LEs $\lambda_n$ ($n\ge2$) smaller than the largest
one $\lambda=\lambda_1$ \cite{Szendro2007,Pazo2008}.
For the Log-BVs the typical plateau size decreases as $\varepsilon_0$ is
increased,
whereas for the sub-dominant LVs the plateau size is known \cite{Szendro2007,Pazo2008}
to decrease as the index $n$ is increased ($\lambda_n\ge\lambda_{n+1}$). 
This suggests an interesting relation
between both (Lyapunov and bred) vector types that is exploited below
in Sec.~\ref{non-leading}.

Note also that Primo {\em et al.}~\cite{Primo2006} used the surface (\ref{surflbv})
to uncover several spatio-temporal scaling relations that already revealed
in an indirect way the existence of cut-off lengths. These cut-off
lengths are now apparent in the light of the plateaus in Fig.~\ref{plateaus}.

\section{bred exponents}
\label{gr}
We now focus on the (exponential) growth rate of Log-BVs, which we will
refer to as bred exponents (BEs).
We define the BEs as:
\begin{equation}
\lambda_{\varepsilon_0} = \frac{1}{\tau} \Big{\langle} \ln \frac{\|{\mathbf
\delta \mathbf u}(\tau_m+ \tau)\|}
{\|{\mathbf \delta \mathbf u}_{\varepsilon_0}(\tau_m)\|}\Big{\rangle},
\label{be1}
\end{equation}
with $\tau$ denoting the time between rescalings.
$\lambda_{\varepsilon_0}$
can be seen as a type of finite-size Lyapunov exponent,
see Appendix A of \cite{Boffetta2002}. 
The value of $\lambda_{\varepsilon_0}$ is not very
sensitive to $\tau$ if $\tau$ remains small.
In our simulations we chose $\tau$ such that the perturbation
does not amplify more than $\exp(1/2)$ times in a breeding cycle,
so we take 
\begin{equation}
\tau \lambda  \lesssim \frac{1}{2},
\end{equation}
with $\lambda$ being the largest LE.
First of all, it is convenient
to transform Eq.~(\ref{be1}), using Eq.~(\ref{defhm}):
\begin{equation}
\lambda_{\varepsilon_0}= \frac{1}{\tau} \left[ \langle
h(x,\tau_m+\tau) \rangle - \ln \varepsilon_0  \right]
\label{be2}
\end{equation}
This expresses in mathematical terms that the BE is the average velocity of the
Log-BV-surface. In the
transformation of (\ref{be1}) into (\ref{be2}) we are assuming the geometric
norm.
Again, this choice makes
plenty of sense because the geometric norm actually yields the least-fluctuating
LE in systems with spatiotemporal chaos
\cite{PikovskyPoliti1998}.

\subsection{Convergence of the Bred Exponent to the first Lyapunov exponent}
\label{sr}

In this subsection we explore the dependence of the
BEs, $\lambda_{\varepsilon_0}(L)$,
on the amplitude $\varepsilon_0$ and on the system size $L$.
Note that in the limit of vanishing amplitude one recovers the
largest LE of the system: $\lambda_{\varepsilon_0\to 0}(L)=\lambda(L)$ 

The key step of our following analysis is the use of the associated surfaces 
and the universal scaling laws they obey.
Let us denote by $\lambda(L=\infty)$ the LE of
the model with infinite size.
As already reported in Ref.~\cite{PikovskyPoliti1998},
the LE of a system of size $L$,
$\lambda(L)$ deviates from the infinite size limit as
\begin{equation}
\lambda(L=\infty) - \lambda(L) \sim \frac{1}{L} 
\label{scaling2}
\end{equation}
This stems from the fact that 
$\lambda(L)$ is the velocity of the associated surface,
that scales as solutions of the KPZ equation:
The asymptotic velocity of a KPZ-surface
presents a system-size correction ~\cite{KM90} of order
$\sim L^{-2(1-\alpha)}$, with $\alpha=1/2$ one dimension.

For the dependence of the BE on $\varepsilon_0$
we resort to very simple arguments.
Following the reasonings in \cite{Primo2005} let us 
assume a Log-BV
is (locally) linear whenever $|\delta u(x,t)| < B$,
for a certain bound $B$.
This bound defines borders of regions of size $l_c$ where the Log-BV-surface
is approximately a copy of the leading LV-surface.
The KPZ surface is self affine, and recalling Eq.~(\ref{defhm}),
we expect the scaling law
\begin{equation}
\ln B-\ln \varepsilon_0 \sim l_c^{\alpha}
\label{eq15}
\end{equation}
to be fulfilled.
Taking into account that the velocity is shifted
from the asymptotic value by the inverse of the
size, {\em i.e.}~$l_c^{-1}$:
\begin{equation}
\lambda(L=\infty)-\lambda_{\varepsilon_0}\sim 
\frac{1}{\left( \ln B - \ln \varepsilon_0\right)^{1/\alpha}} 
\label{scaling1}
\end{equation}
This relation is not the asymptotic one because when $\varepsilon_0$ becomes
extremely small
the ``nonlinear barrier'' at $\ln B$ is
seldom achieved
and the dominant correction is given by Eq.~(\ref{scaling2}).

\begin{figure}
\epsfig{file=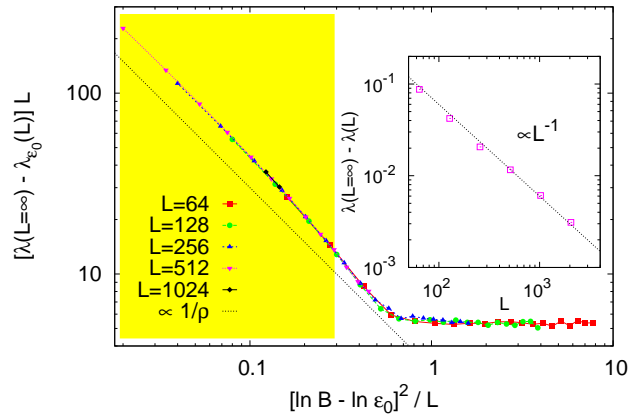,width=6cm, angle= -90}
\caption{\label{figscaling}
(color on line) Data collapse of $\lambda_{\varepsilon_0}(L)$, defined in 
(\ref{be1}) and (\ref{be2}),
for different values of $\varepsilon_0$ and system size $L$ ($F=8$).
Points fall approximately on a line,
the scaling function $g$ in (\ref{scaling}). Inset:
The value of the LE
$\lambda(L=\infty)$ was determined by extrapolating the
largest LE from finite systems using (\ref{scaling2})}
\end{figure}

Scaling relations (\ref{scaling2}) and (\ref{scaling1})
can be cast into
\begin{equation}
\lambda(L=\infty)-\lambda_{\varepsilon_0}(L) = f(\rho)/L
\label{scaling}
\end{equation}
where $\rho=(\ln B - \ln\varepsilon_0)^2/L$ and $f(\rho)$ is a scaling function:
For $\rho \gg 1$, $f(\rho)={\rm const.}$, as expected from (\ref{scaling2});
whereas for $\rho \ll 1$, $f(\rho)$ 
should scale as $\sim \rho^{-1}$, according to (\ref{scaling1}).
Using the data from our simulations we can test the validity
of this scaling law. Figure~\ref{figscaling} shows a very good collapse
with just one fitting parameter, $\ln B=2.2$,
for different system sizes and values of $\varepsilon_0$.
Our scaling argument is in good agreement with numerical data shown in
Fig.~\ref{figscaling}, where one can see how the $\rho \ll 1$ asymptote
converges to $1/\rho$ (shaded region in Fig.~\ref{figscaling}). However, the
approximation appears to be too crude to
describe the behavior very close to the crossover point $\rho \approx 1$, where
the functional form of $f(\rho)$ deviates from the asymptote. 
Nonetheless the
correct scaling variable $\rho$ for the scaling relation (\ref{scaling}) and
the two asymptotes are correctly captured.
The crossover of $f(\rho)$ at $\rho\approx1$ marks
the departure from a regime dominated by the finite-size system effects
to the dominance of the amplitude $\varepsilon_0$ of the Log-BV.

\subsection{Bred exponents vs. Lyapunov spectrum}
\label{non-leading}
In the preceding subsection we have introduced the concept of bred vector
exponent $\lambda_{\varepsilon_0}$ as the growth rate for a Log-BV of amplitude
$\varepsilon_0$. We have also seen how this BE approaches the leading LE in a
finite size system as the amplitude $\varepsilon_0$ is varied. Now we devote
the present subsection to study the connection of the BEs with the
spectrum of LEs. The question we want to address is to what extent growth rates of
Log-BVs (finite errors) approach growth rates of LVs (infinitesimal errors).

\subsubsection{Main hypothesis}
From the scaling arguments discussed above, 
it can be expected that the BE $\lambda_{\varepsilon_0}$  ought to be close
to the $n$-th Lyapunov exponent $\lambda_n$ whenever both of them
are piecewise copies of the leading Lyapunov vector with similar plateau sizes. 
This hypothesis reads more formally:
\begin{equation}
 \lambda_{\varepsilon_0} \approx \lambda_n  \; \; {\rm if} \; \; 
 l_c^{BV}(\varepsilon_0) \approx l_c^{LV}(n) \, ,
\label{hypot}
\end{equation}
where $l_c^{BV}$ ($l_c^{LV}$) indicates the typical length scale over which the
Log-BV (the $n$-th LV)
surface is a piecewise copy of the leading LV.
%
%
In the following we will drop the superindices $BV$ and $LV$, as it is clear from the argument
($\varepsilon_0$ or $n$) what vector type we are referring to.
\begin{figure}
\epsfig{file=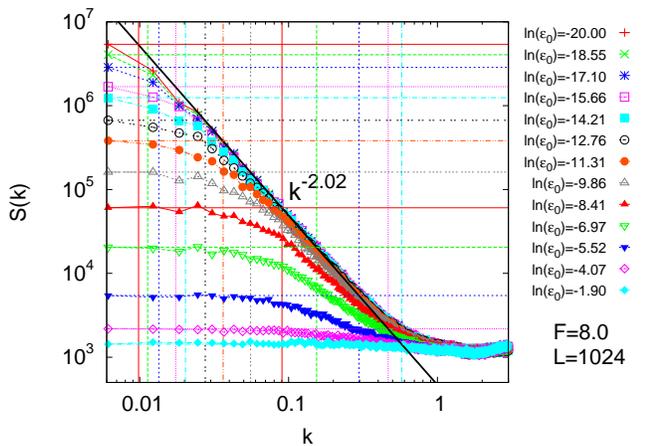,width=6cm, angle= -90}
\caption{\label{strucfacdemo}
(color online) The lines with symbols are structure factors $S(k)$ obtained for
Log-BVs with different perturbation amplitudes for $F=8.0$ and $L=1024$. 
The straight lines without symbols indicate the method of estimating the values
$k_c(\varepsilon_0)$ which
are determined from the intersection of a horizontal line
at $S(k_{min}=2\pi/L)$ and a power-law $S(k) \propto k^{-2.02}$ obtained from
fitting the
structure factor for $\min(\ln \varepsilon_0)$.}
\end{figure}

\subsubsection{Estimation of crossover wavenumbers $k_c(\varepsilon_0)$}
In Sec.~\ref{sgfb} we showed that Log-BVs are piecewise copies of the main LV.
This observation is now exploited to obtain a connection between BEs and LEs.
In order to do so, we find it very convenient to look at the form of spatial
correlations in Fourier space.
The structure factors of Log-BV-surfaces with different values of
$\varepsilon_0$ are presented in Fig.~\ref{strucfacdemo}. 
As we have seen in Fig.~\ref{plateaus} the Log-BV-surface 
follows the leading LV-surface at short scales and, as a consequence,
their respective structure factors should 
overlap above a certain $k_c(\varepsilon_0)$.
This implies that both (Log-BV and LV) surface types share an interval
$k_c(\varepsilon_0)<k<\bar k \simeq 0.3$
with power-law structure factor $S(k)\sim k^{-2}$.
Log-BVs exhibit flat structure factors for $k$ smaller than a certain $k_c(\varepsilon_0)$,
which indicates that distant regions (corresponding to small $k$) are basically
uncorrelated.
The crossover wavenumber $k_c(\varepsilon_0) \sim l_c(\varepsilon_0)^{-1}$ is
monotonically increasing with $\varepsilon_0$ and,
from Eq.~(\ref{eq15}), we have $k_c(\varepsilon_0) \sim (\ln B - \ln
\varepsilon_0)^{-1/\alpha}$ with $\alpha =1/2$.
Notably, surfaces associated with LVs for $n>1$ have a 
structure factor that also exhibits a knee at a certain crossover wavenumber $k_c(n)$
\cite{Szendro2007,Pazo2008}. 
In sum, structure factors of LV-surfaces and Log-BV-surfaces are
very similar (cf.~Fig.~4 in \cite{Pazo2010}),
and they only differ in the algebraic dependence below the particular
crossover wavenumber $k_c$: $1/k$-type for the LVs, and flat ($k^0$) 
for the Log-BVs \cite{Pazo2010}, reflecting their different spatial correlations
at large scales.
\begin{figure}
\epsfig{file=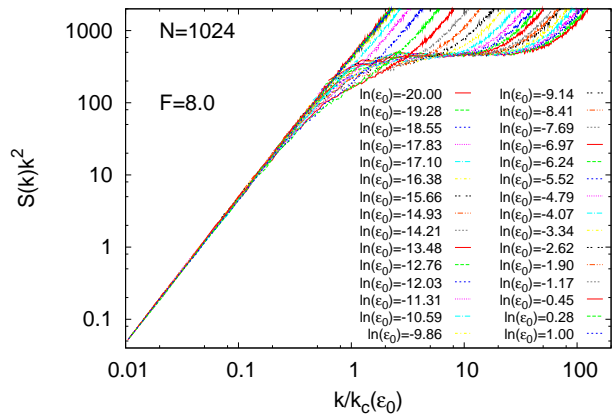,width=6cm, angle= -90}
\caption{\label{collapse} (color online)
Collapse of the structure factors using the 
the values $k_c(\varepsilon_0)$ determined from the structure factors $S(k)$
in Fig.~\ref{strucfacdemo}.}
\end{figure}

\begin{figure}
\epsfig{file=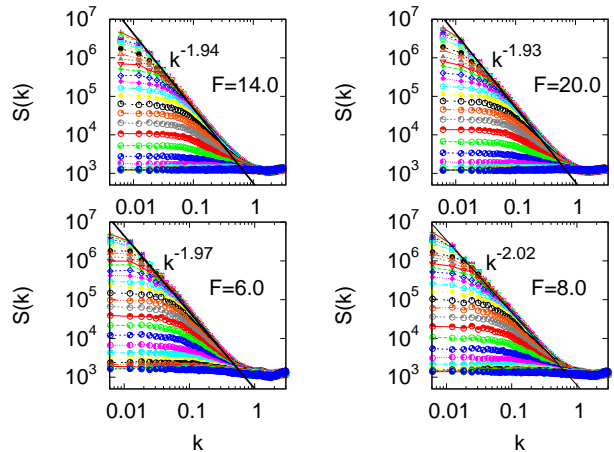,width=6cm, angle= -90}
\caption{\label{sofk1024}
The lines with symbols indicate structure factors $S(k)$ obtained for
bred vectors with different perturbation amplitudes with $F=6,8,14,20$ and $L=1024$. %
The values $k_c(\varepsilon_0)$ are determined like in Fig.~\ref{strucfacdemo}.}
\end{figure}

The crossover wavenumbers $k_c(\varepsilon_0)$ of the
Log-BV-surfaces $h_{\varepsilon_0}(x,t)$
can be systematically extracted from their structure factors $S(k)$,
see Fig.~\ref{strucfacdemo} for $L=1024$ and $F=8$. Figure \ref{collapse} shows the
collapse of different structure factors when normalizing the wavenumbers
by $k_c(\varepsilon_0)$, giving an idea of the goodness of the estimation of $k_c(\varepsilon_0)$.
Figures \ref{sofk1024} and \ref{sofk128} show that the same procedure can be used for other
values of $F$ and smaller systems.

\subsubsection{Numerical results}
\begin{figure}
\epsfig{file=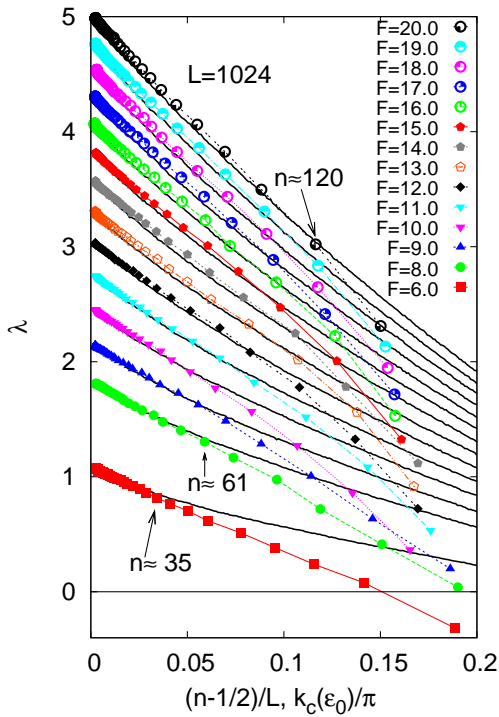, width=7cm}
\caption{\label{bexp1024}
(color online) Lyapunov spectra (solid black lines) 
and bred exponents (symbols).
Size $L=1024$, and $F=6,7,\ldots, 20$.}
\end{figure}

In extended systems, it is customary to represent LEs with the index
$n$ normalized by the system size (or by the number of degrees of freedom).
This is done so to highlight the extensivity (assuming it exists) of the model
because LEs for different systems sizes with identical parameter values 
will approximately fall on the same line.
Figure \ref{bexp1024} shows the LEs connected by black lines.
The Lyapunov spectra were obtained
using the standard algorithm by
Benettin  {\em et al.}~\cite{Benettin1980-1, Benettin1980-2, Wolf1985}.

We found in Ref.~\cite{Szendro2007} for an extensive coupled map lattice that
$k_c(n)\sim(n/L)^\theta$,
with $\theta$ around one. This means that representing the LEs versus $(n/L)$ is,
for moderate values of $n$, equivalent to use a quantity proportional to $k_c(n)$ in the $x$-axis.

If now we want to test our hypothesis that BEs and LEs are similar if the
crossover length scales coincide, we have to normalize
$k_c(\varepsilon_0)$ to use it as an independent variable in the range $(0,1)$. 
This normalization yields $k_c(\varepsilon_0)/\pi$, and we may see in
Fig.~\ref{bexp1024} that indeed BEs and LEs fall very near in
the first part of the Lyapunov spectrum.
This range of approximate overlapping is limited by the value of $n$ (or $\varepsilon_0$) up to
which the LVs (or the Log-BVs) are piecewise copies of the main LV. 
Note also that this range expands as
$F$ increases and the system becomes more chaotic.

\begin{figure}
\epsfig{file=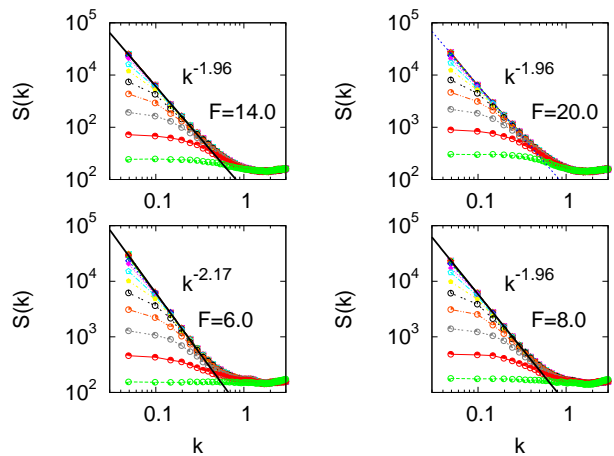,width=6cm, angle= -90}
\caption{\label{sofk128}
(color online) The colored lines with symbols indicate structure factors $S(k)$ obtained for
Log-BVs with different perturbation amplitudes for $F=6,8,14,20$ and $L=128$. 
The values $k_c(\varepsilon_0)$ are determined like in Fig.~\ref{strucfacdemo}
}
\end{figure}

\begin{figure}
\epsfig{file=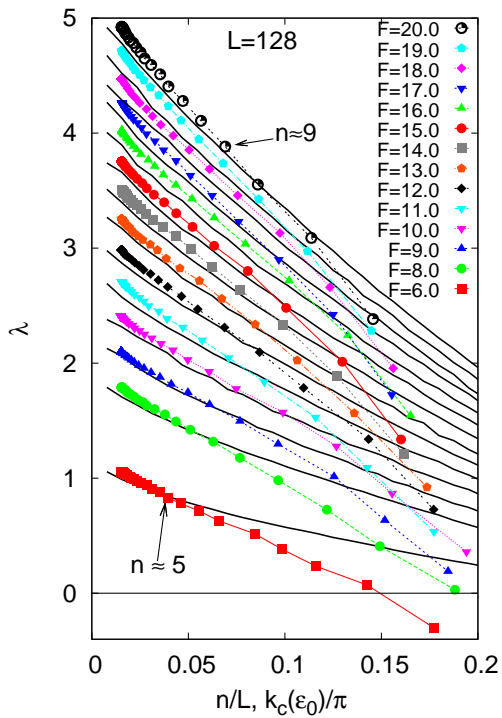,width=7cm}
\caption{\label{bexp128}
(color online) Lyapunov spectra (solid black lines) 
and bred exponents (symbols).
Size $L=128$, and $F=6,7,\ldots, 20$.}
\end{figure}

Figures~\ref{sofk128} and \ref{bexp128} show that the same procedure can be
followed also for a smaller system size ($L=128$). Note that as the system is extensive
the figures look very much the same as those for $L=1024$.

\section{Conclusions}
\label{conc}
Although dynamical systems theory can characterize essentially all the
properties of a chaotic system in terms of the properties of the tangent 
space
directions (LVs) and their associated growth rates, the use of these linear
analysis tools in real applications is often restricted by practical limits.
This is particularly obvious in atmospheric dynamics and weather
prediction systems were operative models include mathematical subtleties 
like
ad-hoc parametrizations of many physical processes, nondifferentiable
structures, and discrete programming structures that render the model 
unsuitable
for linearization.
In this context, much insight has been achieved by studying finite amplitude and
truly nonlinear perturbations. 
In particular
BVs have attracted much attention as a tool to
investigate propagation of errors in both toy and operative weather models.
However, little was known about the relation of finite BVs and
truly infinitesimal LVs, apart from the obvious fact that a BV should 
tend to
be collinear with the leading LV as the BV amplitude $\varepsilon_0$ tends 
to zero (independently of the norm definition used).

In this paper we have uncovered a number of
connections between bred and Lyapunov vectors, which are
specially noticeable and most conveniently characterized when one uses the
zero-norm BVs or Log-BVs. We have also found that these similarities 
appear over
a spatial range, below some characteristic length scale, that depends on
the BVs amplitude.
After a Hopf-Cole transformation, the
Log-BVs turn out to be a piecewise copy of the leading LV.
This resembles
what has been previously reported for non-leading
LVs~\cite{Szendro2007,Pazo2008}. Interestingly,
the spatial structure of Log-BVs shows clearly that they are
uncorrelated objects at long length scales over certain characteristic 
length. This
contrasts with LVs that were shown to exhibit
weak correlations (decaying as $1/k$ with the wavenumber) at long
scales~\cite{Szendro2007,Pazo2008}. This immediately
implies that the relative heights of
the plateaus relating Log-BVs and the LV are not constrained by the dynamics.
In turn different initial conditions
yield a diversity of Log-BVs but nonetheless similar local patterns.
Interestingly, this ``regional''  coincidence was observed before in 
bred vectors
of a global circulation model of the atmosphere \cite{TothKalnay1997},
but remained up to now unexplained. Here we have shown that this regional
similarities of BVs of different amplitudes $\varepsilon_0$ may be understood
as due to the overlapping with the leading LV below a certain spatial range.

This paper has also investigated the relation between
the growth rate of Log-BVs (the BEs)
and the LEs.
We have 
introduced the
concept of bred exponents that describe how nonlinear perturbations grow in
time. We have found that BEs and LEs can be mapped onto each other when
the crossover length scales of the corresponding vector perturbations 
coincide.
This is only true for the most expanding (bred or Lyapunov) exponents,
when the ``piecewise KPZ'' picture holds.
The convergence of the BE (a finite-time LE) to the
largest LE has been found to follow a generic scaling function,
which has been explained by a simple argument.

Finally, it is to be emphasized that we have used the Lorenz '96 model 
in our
simulations,
but very similar results should be expected for other
dissipative systems with spatiotemporal chaos.
\acknowledgments
Financial support from the Ministerio de Ciencia e Innovaci\'on (Spain) under projects
FIS2009-12964-C05-05 and CGL2007-64387/CLI is acknowledged.
D.P.~acknowledges support by CSIC under the Junta de Ampliaci\'on de
Estudios Programme (JAE-Doc).


\bibliographystyle{prsty}

\begin{thebibliography}{10}

\bibitem{TothKalnay1993}
Z. Toth and E. Kalnay, Bull. Amer. Meteor. Soc. {\bf 74},  2317  (1993).

\bibitem{kalnay}
E. Kalnay, {\em Atmodpheric {M}odelling, {D}ata {A}ssimilation and
  {P}redictability} (Cambridge University Press, Cambridge, 2003).

\bibitem{tlm}
S. Polavarapu, M. Tanguay, R. M\'enard, and A. Staniforth, Tellus {\bf 48A},
  74  (1996).

\bibitem{Primo2005}
C. Primo, M.~A. Rodr{\'{\i}}guez, J.~M. Lop{\'e}z, and I.~G. Szendro, Phys.
  Rev. E {\bf 72},  015201  (2005).

\bibitem{Primo2006}
C. Primo, M.~A. Szendro, I.~G.~and~Rodr{\'i}guez, and J.~M. L{\'o}pez,
  Europhys. Lett. {\bf 76},  767  (2006).

\bibitem{Boffetta2002}
G. Boffetta, M. Cencini, F. M., and A. Vulpiani, Phys. Rep. {\bf 356},  367
  (2002).

\bibitem{Lorenz1996}
E.~N. Lorenz,  in {\em Proc. Seminar on Predictability Vol. I}, ECWF Seminar,
  edited by T. Palmer (ECMWF, Reading, UK, 1996), pp.\ 1--18.

\bibitem{Lorenz1998}
E.~N. Lorenz and K.~A. Emanuel, J. Atmos. Sci. {\bf 55},  399  (1998).

\bibitem{Lorenz2006}
E.~N. Lorenz, J. Atmos. Sci. {\bf 63},  2056  (2006).

\bibitem{Karimi2009}
A. Karimi and M.~R. Paul, arXiv:0906.349v1 [nlin.PS] (unpublished).

\bibitem{CrossHohenberg}
M.~C. Cross and P.~C. Hohenberg, Rev. Mod. Phys. {\bf 65},  851  (1993).

\bibitem{Pazo2010}
D. Paz{\'o}, M.~A. Rodr\'{\i}guez, and J.~M. L\'opez, Tellus {\bf 62A},  10
  (2010).

\bibitem{GiacomelliPoliti1991}
G. Giacomelli and A. Politi, Europhys. Lett. {\bf 15},  387  (1991).

\bibitem{Pikovsky1994}
A. Pikovsky and J. Kurths, Phys Rev. E {\bf 49},  898  (1994).

\bibitem{PikovskyPoliti1998}
A. Pikovsky and A. Politi, Nonlinearity {\bf 11},  1049  (1998).

\bibitem{Pazo2008}
D. Paz{\'o}, I.~G. Szendro, J.~M. L{\'o}pez, and M.~A. Rodr{\'i}guez, Phys Rev.
  E {\bf 78},  016209  (2008).

\bibitem{Kardar1986}
M. Kardar, G. Parisi, and Y.~C. Zhang, PRL {\bf 56},  889  (1986).

\bibitem{Barabasi1995}
A.~L. Barab{\'a}si and H.~E. Stanley, {\em Fractal Concepts in Surface Growth}
  (Cambridge University Press, Cambridge, 1995).

\bibitem{Halpin-Healy1995}
T. Halpin-Healy and Y.-C. Zhang, Phys. Rep. {\bf 254},  215  (1995).

\bibitem{Szendro2007}
I.~G. Szendro, D. Paz{\'o}, M.~A. Rodr{\'i}guez, and J.~M. L{\'o}pez, Phys Rev.
  E {\bf 76},  025202R  (2007).

\bibitem{KM90}
J. Krug and P. Meakin, J. Phys. A: Math. Gen. {\bf 23},  L987  (1990).

\bibitem{Benettin1980-1}
G. Benettin, L. Galgani, A. Giorgilli, and J.~M. Strelcyn, Meccanica {\bf 15},
  9  (1980).

\bibitem{Benettin1980-2}
G. Benettin, L. Galgani, A. Giorgilli, and J.~M. Strelcyn, Meccanica {\bf 15},
  21  (1980).

\bibitem{Wolf1985}
A. Wolf, J.~B. Swift, H.~L. Swinney, and J.~A. Vastano, Physica D {\bf 16},
  285  (1985).

\bibitem{TothKalnay1997}
Z. Toth and E. Kalnay, Mon. Wea. Rev. {\bf 125},  3297  (1997).

\end{thebibliography}

\end{document}